\documentclass[aps,prd,superscriptaddress,nofootinbib,floatfix,preprint]{revtex4-1}

\usepackage[dvipdfmx]{graphicx}
\usepackage{amsmath}
\usepackage{textcomp}
\usepackage{mathcomp}
\usepackage{bm}
\usepackage{here}
\usepackage[small,bf]{caption}
\usepackage[subrefformat=parens]{subcaption}
\usepackage{comment}

\newcommand{\beq}{\begin{equation}}
\newcommand{\eeq}{\end{equation}}
\newcommand{\beqa}{\begin{eqnarray}}
\newcommand{\eeqa}{\end{eqnarray}}
\newcommand{\bpr}{\begin{problem}}
\newcommand{\epr}{\end{problem}}
\newcommand{\bcent}{\begin{center}}
\newcommand{\ecent}{\end{center}}
\newcommand{\bfig}{\begin{figure}}
\newcommand{\efig}{\end{figure}}
\newcommand{\bpc}{\begin{picture}}
\newcommand{\epc}{\end{picture}}
\newcommand{\barr}{\begin{array}}
\newcommand{\earr}{\end{array}}
\newcommand{\bitm}{\begin{itemize}}
\newcommand{\eitm}{\end{itemize}}
\newcommand{\bright}{\begin{flushright}}
\newcommand{\eright}{\end{flushright}}
\newcommand{\bminip}{\begin{minipage}}
\newcommand{\eminip}{\end{minipage}}
\newcommand{\btab}{\begin{tabular}}
\newcommand{\etab}{\end{tabular}}

\newcommand{\hiroshima}{Graduate School of Advanced Science and Engineering, Hiroshima University, Kagamiyama, Higashi-Hiroshima, Hiroshima 739-8526, Japan}

\newcommand{\tSRPC}{$\mathrm{^tSRPC}$}

\begin{document}
\title{
Design and construction of a variable-angle three-beam stimulated resonant photon collider
toward eV-scale ALP search
}

\author{Takumi Hasada}\affiliation{\hiroshima}
\author{Kensuke Homma\footnote{corresponding author}}\affiliation{\hiroshima}
\author{Yuri Kirita}\affiliation{\hiroshima}

\date{\today}

\begin{abstract}
We aim at search for axion-like particles in the eV mass range using a variable-angle
stimulated resonance photon collider (SRPC) with three intense laser beams. 
By changing angle of incidence of the three beams, the center-of-mass-system
collision energy can be varied and the eV mass range can be continuously searched for. 
In this paper, we present the design and construction of such a 
variable-angle three-beam SRPC (\tSRPC),
the verification of the variable-angle mechanism using a calibration laser,
and realistic sensitivity projections for the near future searches.
\end{abstract}

\maketitle

\section{Introduction}
Some of the unsolved problems of the Standard Model may be answered by new particles in the low-mass region, 
which has not yet been fully explored. 
Nambu-Goldstone bosons (NGBs) which are supposed to be ideally massless may appear whenever
global continuous symmetries spontaneously break. 
Axion is a representative candidate for such new fields in the low-mass regime
because axion is a pseudo Nambu-Goldstone boson (pNGB)
arising from spontaneous breaking of the Peccei-Quinn symmetry \cite{PQ}
introduced to solve the strong CP problem\cite{CP} in the context of Quantum ChromoDynamics (QCD).
More generalized low-mass particles are called Axion-Like-Particles (ALPs), some of which 
can also be reasonable dark matter candidates.

Many experiments have been conducted to detect ALPs focusing on their coupling to photons 
given by the following interaction Lagrangian
$\mathcal{L}= -\frac{1}{4} \frac{g}{M} F_{\mu \nu} \tilde{F}^{\mu \nu} a$, 
where $a$ is an ALP field, $F_{\mu \nu}$ is the electric field strength, 
$\tilde{F}^{\mu \nu}$ is its dual, and $g/M$ is the coupling constant with dimensionless parameter $g$
and $M$ denoting an energy scale at which a symmetry breaking takes place. 
From the experimental point of view, a mass range between 0.001 and 10 eV has not been intensively explored
especially by laboratory-based experiments.
In addition to the QCD axion scenario,
a model {\it miracle} predicts the existence of an ALP in the mass range 0.01$\sim$1eV as a possible explanation 
for both inflation and dark matter~\cite{miracle}.
It is thus very intriguing to conduct exploratory experiments 
using laser fields in the near-infrared region to have sensitivities to the eV range. 

We have been performing the ALP search based on 
stimulated resonant photon colliders (SRPC) concept~\cite{Fuji,JHEP2020,PTEP2014}. 
In this method, a single pulsed creation laser are focused and arbitrary two photons included in the field 
collide with each other resulting in production of an ALP and another pulsed inducing laser simultaneously
stimulates its decay. 
This method does not require any assumptions except the interaction Lagrangian, thus,
independent of any cosmological and astrophysical models.
In our previous study of quasi-parallel collision system,
we focused two lasers into the same optical axis 
and collided them at a shallow incident angle to search specifically 
for sub-eV masses~\cite{PTEP2014,PTEP2015,PTEP2020,SAPPHIRES00,SAPPHIRES01}. 
Recently, we have proposed and demonstrated a pilot ALP search based on 
three-beam stimulated resonant collisions (\tSRPC) for heavier masse range\cite{3beam00,3beam01}. 
In the near future, we plan to search for ALPs by
continuously changing their collision angles in the eV mass range.

In this paper, we present the design and construction of a three beam SRPC that 
can continuously scan the eV mass range by changing the incident angles of the three colliding lasers. 
We then show the verification of the mechanism of 
collision angle changes for individual mass range using a He:Neon laser for the calibration purpose. 
In the following sections, we first discuss a choice of the basic design to introduce the variable collision angles.
Secondly, we introduce the concrete design for the variable-angle stimulated resonant photon collider
by taking several aspects of calibration steps into account.
Thirdly, we provide the verification of the selected mechanism using the calibration laser. 
We finally discuss the achievable sensitivity projections based on a realistic experimental parameter set
and conclude the paper.

\section{Kinematics in three-beam Stimulated Resonant Photon Collider, \tSRPC}
\begin{figure}[]
\begin{center}
\includegraphics[keepaspectratio,scale=0.7]{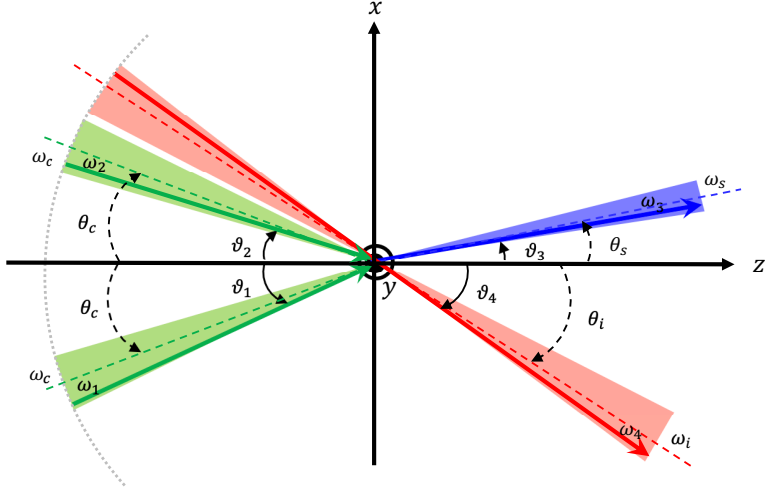}
\caption{
Concept of a three-beam stimulated resonant photon collider (\tSRPC) 
with focused coherent beams~\cite{3beam01}. 
The two focused creation laser beams (green) at the incident angle $\theta_{c}$
produces an ALP resonance state and the focused inducing laser beam (red) stimulates its decay.
The creation photons have different energies $\omega_{1}$ and $\omega_{2}$ from the central
value of $\omega_{c}$ and 
different incident angles $\vartheta_{1}$ and $\vartheta_{2}$ from $\theta_{c}$ respectively. 
Similarly, the inducing laser (red) with a central wavelength of $\omega_{i}$ has 
part of the beam with $\omega_{4}$, increasing the emission probability of 
the signal photon of $\omega_{3}$ (blue) via energy-momentum conservation.
}
\label{Fig1}
\end{center}
\end{figure}
Figure \ref{Fig1} shows the conceptual drawing of \tSRPC. 
We consider a case that two creation laser pulses (green) are
focused at the same incident angle $\theta_{c}$ with 
the same energy $\omega_{c}$ and similarly an inducing laser pulse (red) is
focused with the energy $\omega_{i}$ which
increases the interaction rate of stimulated scattering 
emitting signal photons of energy $\omega_{s}$ (blue) that satisfies energy-momentum conservation. 
Energy-momentum conservation between three beams photons and a signal photon
requires the following kinematical relations:
\begin{equation}\label{eq1}
    \begin{aligned}
\omega_{c}+\omega_{c} &= \omega_{s}+\omega_{i} \\
2 \omega_{c} \cos \theta_{c} &= \omega_{s} \cos \theta_{s}+\omega_{i} \cos \theta_{i} \\
\omega_{s} \sin \theta_{s} &= \omega_{i} \sin \theta_{i}.
    \end{aligned}
\end{equation}
Experimentally we first fix beam energies $\omega_c$ and $\omega_i$ among available laser wavelengths.
We then target a ALP mass of $m_a$ which coincides with the center of mass system energy $E_{cms}$
defined as
\begin{equation}\label{eq2}
m_a = E_{cms} = 2\omega_c \sin\theta_c.
\end{equation}
From Eqs.(\ref{eq1}) and (\ref{eq2}) the angle of incidence for the inducing beam 
can be determined as follows
\begin{align}\label{eq3}
\theta_{i} &= \arccos\left\{ \left(1-\frac{m_{a}^2}{4 \omega_{c} \omega_{i}}\right)\left(1-\frac{m_{a}^2}{4 \omega_{c}^2}\right)^{-\frac{1}{2}} \right\}.
\end{align}

We emphasize, however, that individual photons within focused beams indeed have different energies 
$\omega_{1}$ and $\omega_{2}$ from $\omega_{c}$ and $\omega_4$ from $\omega_{i}$.
These energy uncertainties are caused by the Fourier transform-limited short-pulse laser. 
In addition in the focused fields,
the incident angles $\vartheta_{1}$, $\vartheta_{2}$, and $\vartheta_{4}$ are 
also different from $\theta_{c}$ and $\theta_{i}$, respectively.
Fluctuations in the angle of incidence around the beam axis are caused by momentum fluctuations near the focal point.
Fortunately, these uncertainties give the center-of-mass collision energy $E_{cms}$ a finite width
via the following relation
\begin{equation}\label{eq4}
E_{cms} = 2\sqrt{\omega_1 \omega_2}\sin\left(\frac{\vartheta_1+\vartheta_2}{2}\right).
\end{equation}
Thus ALP mass scanning is possible even though central values $\theta_{c}$ are varied in discrete steps
if the $E_{cms}$ uncertainty defined by laser pulse duration and the focal parameters
is consistent with the discrete angle step in $\theta_c$ in a search. 
For more information, see \cite{3beam00}.

\section{Basic design to realize variable collision angles}\label{sec2}
\begin{figure}[H]
\begin{center}
\includegraphics[keepaspectratio,width=150mm]{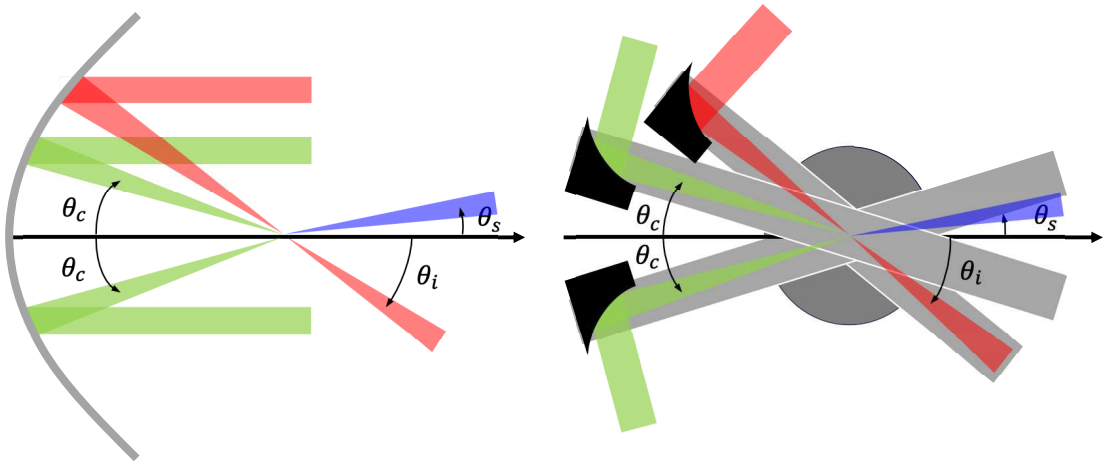}
\caption{Two proposals for variable angle mechanisms. Left:parabolic mirror type
where the collision angle is changable by changing the incident position of lasers. 
Right:rotating stage type where the collision angle is changable 
by assembling a beam focusing system on multiple rotary stages.}
\label{Fig2}
\end{center}
\end{figure}

In order to realize continuously changable collision angles between focused three beams, 
the following two main ideas were considered. 
Figure \ref{Fig2} (left) shows a natural idea to change collision angles by changing incident positions
of lasers on a parabolic mirror surface, while Fig.\ref{Fig2} (right) shows a focusing system
on multi-layered rotating stages where angles of incidence in the individual layers are changable 
by independently rotating the individual stages. 

\begin{table}[]
  \caption{Comparison of variable-angle mechanisms.}
  \centering
  \begin{tabular}{lcc}
    \hline
    Item  & Parabolic mirror type  &  Rotary stage type  \\
    \hline \hline
    Adjustment   & easy   & difficult \\
    Size         & large  & compact (vacuum chamber compatible) \\
    Angle range  & narrow & wide \\
    Focal length & angle-dependently variable  &  fixed \\
    Flexibility  & low (custom-ordered mirror)  &  high (catalog items) \\
    \hline
  \end{tabular}
\label{Tab1}
\end{table}
Table \ref{Tab1} shows comparison between two ideas
based on the advantages and disadvantages.
While the parabolic mirror type has advantages of fewer optical components and 
readiness of angle adjustment, 
it is necessary to prepare a custom-made large-area mirror to reach heavier mass region 
with large incident angles $\theta_c$ and $\theta_{i}$. 
The search must be performed in a vacuum chamber in order to reduce the four-wave mixing (FWM) 
originating from ambient atoms. 
Therefore, a large parabolic mirror is not suitable for implementing it in the vacuum chamber.
We also note that the focal length must slightly change for individual incident angles.

On the other hand, the rotary stage type allows scanning over a wide range of incident angles 
in a compact size. 
In addition, it can be constructed using a combination of commercial products, hence, 
has the advantage of flexibility for the design so that we can replace focusing mirrors as we need. 
As we demonstrated in the previous pilot search~\cite{3beam01},
we indeed found that it was necessary to prepare a space for a target holder at the focal point 
to ensure the spatiotemporal synchronization of three pulses, 
a camera system to record the beam profile, and a shielding wall to suppress background from beam remnants. 
This requires lots of flexibility including changing the focal length, which has impact on the sensitivity.
However, there is a disadvantage of the rotary stage type in narrow incident angles because focusing 
optical elements spatially overlap between two incident creation beams.

Therefore, in this study, we adopted the hybrid concept combining good futures
of the parabolic mirror type and the rotary stage type depending on incident angles of creation lasers.
For large incident angles, the design is based on the rotary stage type with two moving stages for individual
creation beams while the inducing beam is fixed at an optical table (LA collider). 
For narrow incident angles, on the other hand, one of the two creation beams and the inducing beam 
share a common parabolic focusing mirror and the mirror is fixed at the optical table
while the remaining one creation beam rotates together with a moving stage (NA collider).
In the next section we discuss the relation between LA and NA setups in detail.

\begin{figure}[]
\begin{center}
\includegraphics[keepaspectratio,scale=0.7]{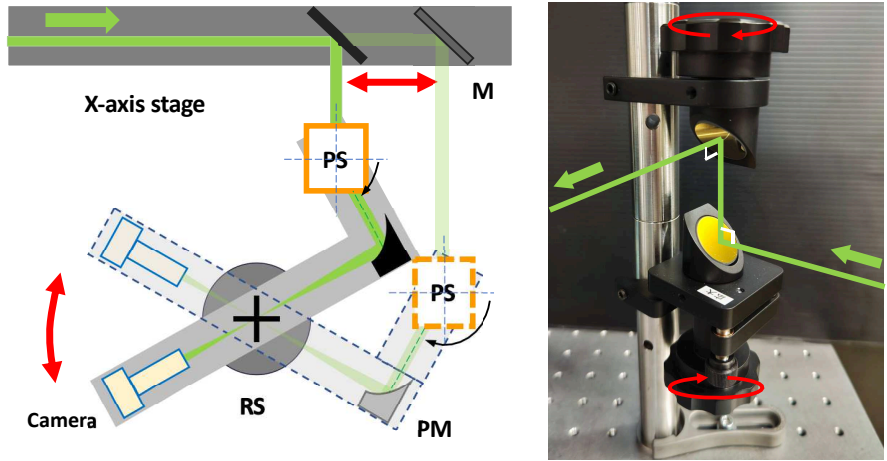}
\caption{Variable angle mechanism using a rotary stage. Incident angle is varied 
by rotating a stage assembling a beam focusing system with a periscope (PS) and a parabolic mirror (PM).
By using a periscope (PS), the angle is changable 
only by rotating the periscope (PS) and the mirror (M) on the x-axis rail stage 
in front of PS. 
By setting the parabolic mirror's focal point at the center of the rotary stage (RS), 
the collision point remains fixed even when the incident angle is varied. 
The focal spots can be checked using a monitoring camera.} 
\label{Fig3}
\end{center}
\end{figure}

The variable-angle mechanism using the rotary stages associates additional complication. 
The rotary stages consist of individual aluminum plates placed on individual rotating stages 
on which a periscope (PS) and a parabolic mirror (PM) to focus a beam are assembled. 
The incident angles are changable by rotating the stage. 
Changing incident angles must accompany change of incident points of lasers.
However, typically the laser incident point is not readily movable 
because high-intensity laser systems are not compact.
In order to compensate this immobility, we introduce periscope (PS) components 
as shown in the picture of Fig.\ref{Fig3}.
PS consists of a pair of mirrors aligned vertically with angle of incidence(AOI) of 45\textdegree. 
PS can bend a beam in any directions by rotating the direction of the mirror in the upside of the PS.
The height of an optical axis is changable by adjusting the relative distance between two mirrors
in the incident and outgoing sides. The change of incident angle is compensated by
the parallel movement of the mirror (M) on the x-axis stage in advance of injection to the PS.
However, since the PS reflects a beam to PM with a possibly large angle, 
it is necessary to evaluate the effect that the incident linearly polarized state 
becomes elliptically polarized one.
As we demonstrated in the pilot search, the change in polarization can be evaluated 
based on the measurement of Stokes parameters~\cite{3beam01}. 
In this collision system, incident angles are changable without moving the collision point 
by setting the PM's focal point at the common center of the rotating stages (RSs). 
The focal spots can be checked using a camera which also rotates around the center.
By stacking layers consisting of an aluminum plate and a rotating stage, 
the collision angle of the three beams can be independently varied. 
We note, however, that it is not necessary to change the angle of all the three colliding beams.
As we discuss later, one of the three beams can be fixed and we have only to adjust AOIs 
of the other two laser beams relative to the fixed one. 

\section{Concrete designs for the large angle and narrow angle setups}\label{sec3}
\begin{figure}[H]
\begin{center}
\includegraphics[keepaspectratio,scale=0.5]{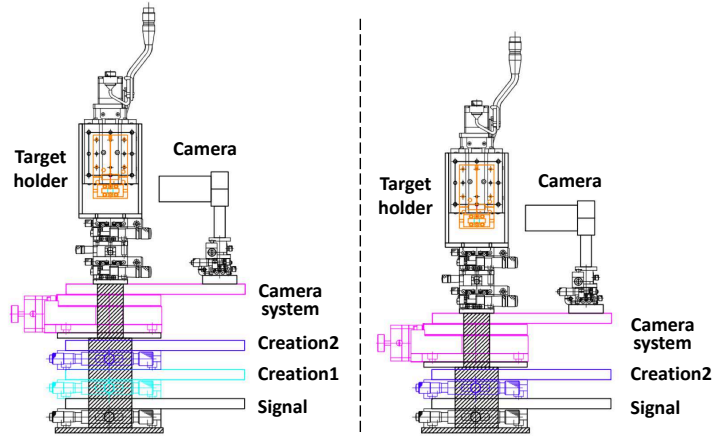}
\caption{
Side views of designed variable-angle three-beam stimulated resonant photon colliders
for large angle (left) and narrow angle (right) setups. 
The detailed explanations are found in the main text.
}
\label{Fig4}
\end{center}
\end{figure}
Figure \ref{Fig4} shows a side view of the designed collision system. 
First, the three beam interaction point (IP) is defined with a thin crossed wire.
The wire target must be immovable with respect to the multi-layer stage movements as discussed below.
The common layers between the LA and NA collision cases are
the bottom signal sampling layer and the top camera layer to monitor beam profiles at IP.
The signal layer is necessary to move the detection
point of generated signal photons because the signal direction must change depending on collision angles.
We introduce two layers for the two creation laser beams in the LA collision case, 
while only one layer for one of the two creation laser
beams is necessary in the NA collision case
because the other creation and the inducing beams share a common parabolic mirror 
which is fixed to the optical table.
The breakdown of the individual layers from the bottom are thus as follows:
signal light (black), creation light 1 (cyan), creation light 2 (purple), and camera system (magenta) for LA
and
signal light (black), creation light 2 (purple), and camera system (magenta) for NA. 

At IP a tower-like spacer is placed to introduce an immovable target which must be independent of movements of 
the four or three rotating stages. The target consists of three components with different positions
aligned along a common vertical line: a thin crossed wires to calibrate space overlap between the three beams, 
a nonlinear optical crystal, BBO, to calibrate synchronization of three laser pulses,
and empty hole to perform the search in vacuum.
The target holder is attached to an automated stage that can move vertically to select the three components
depending on purposes.
The dynamic range of incident angles is determined by the geometric limitation of the optical elements. 
At shallow collision angles, parabolic mirrors start interfering with each other, 
while at wide collision angles, aluminum layers start interfering with each other. 

\begin{figure}[]
 \begin{minipage}[t]{1.0\linewidth}
  \centering
  \subcaption{ }\label{Fig5a}
  \includegraphics[keepaspectratio,scale=0.7]{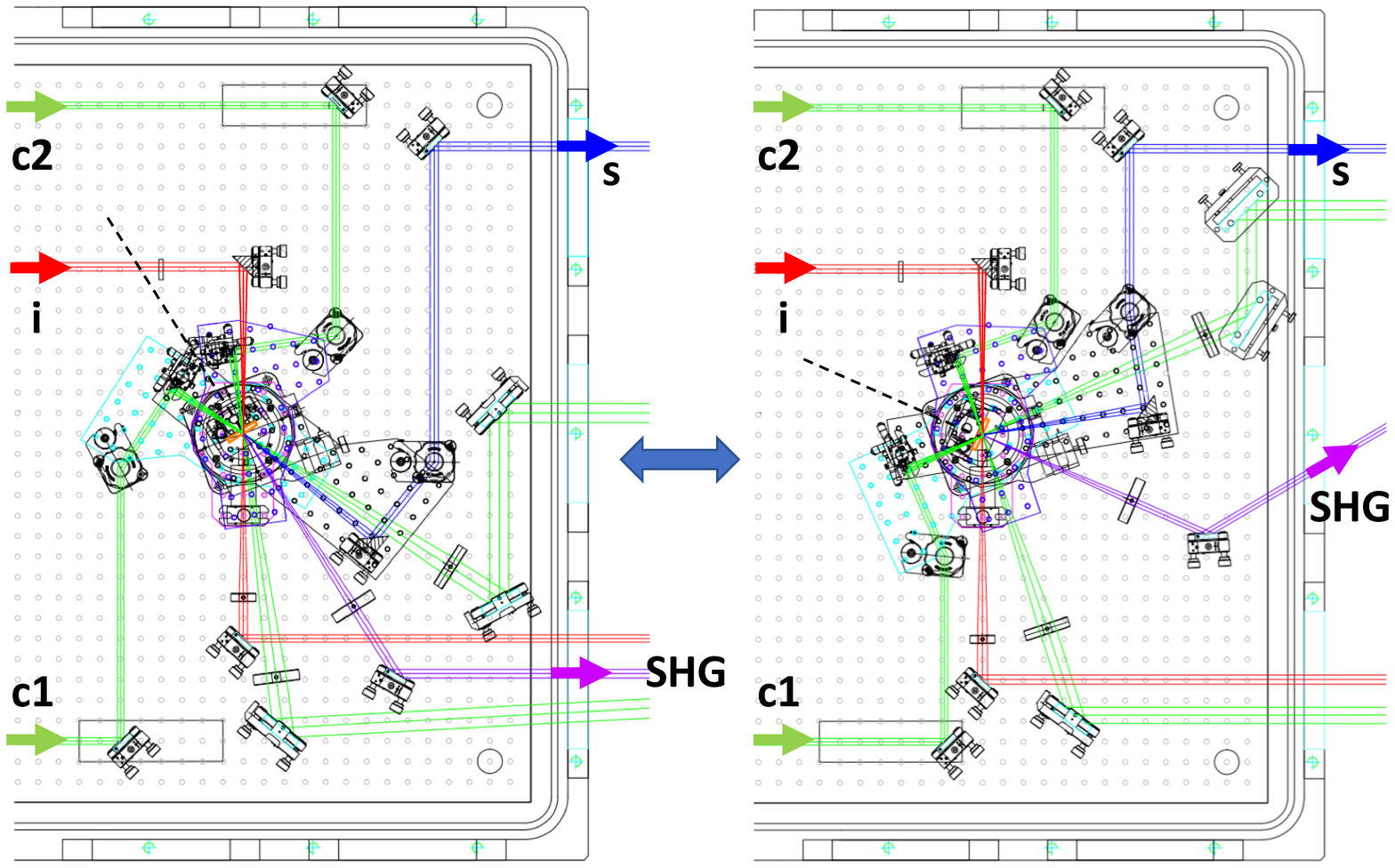}
 \end{minipage} \\
 \begin{minipage}[t]{1.0\linewidth}
  \centering
  \subcaption{ }\label{Fig5b}
  \includegraphics[keepaspectratio,scale=0.8]{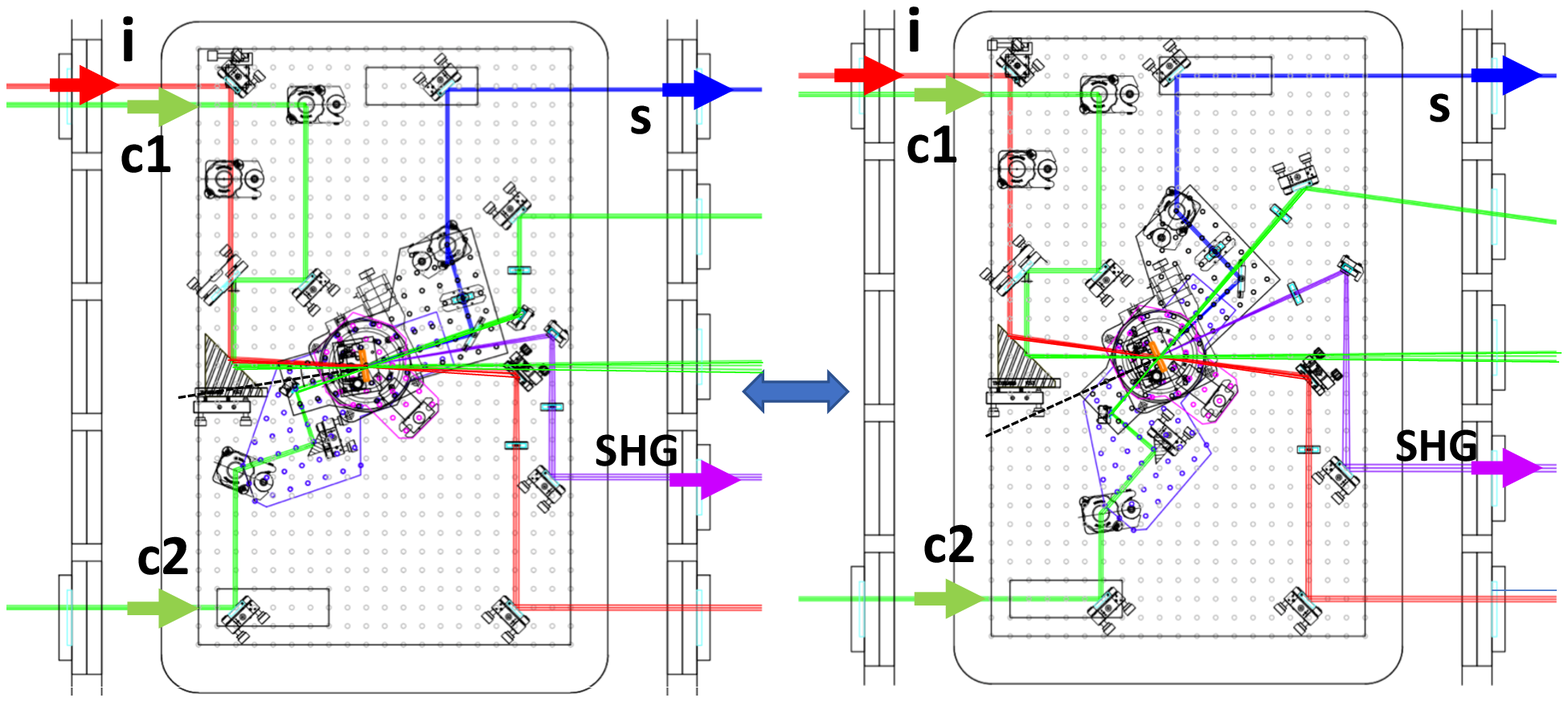}
 \end{minipage} \\
\caption{
The collisional geometries viewed from top. 
(a) large angle setup from $\theta_c=24.8$~deg (left) to $\theta_c=47.9$~deg (right) and (b) narrow angle setup
from $\theta_c=9.3$~deg (left) to $\theta_c=24.8$~deg (right).
The details are explained in the main text.
}
\end{figure}

Figure \ref{Fig5a} shows the rotary-stage-type geometries in the LA setup covering
a large angle range from 24.8 deg.(left) to 47.9 deg.(right). 
The search eventually must be conducted in a vacuum chamber to suppress the atomic background processes. 
Therefore, we aim at a compact design that can be housed in a vacuum chamber. 
The incoming laser from the left represents the creation beam 1, creation beam 2, 
and inducing beam (c1, c2, i), respectively. 
The beam-like objects after the focal point represent the second harmonic generation (SHG) 
and the four-wave mixing photons corresponding signal photons (s). 
In order to guide the generated signal photons to a sensor (photomultiplier, PMT), 
a calibration laser mimicking signal photon trajectories and the wavelength is necessary.
The red inducing beam can be fixed at the bottom optical table in the vacuum chamber,
while the incident angles of the other beams can be aligned relative to this beam. 
The incident angles are changable by rotating creation stages, which requires re-adjustment of PS and M
on the one-axis stage as illustrated in Fig.\ref{Fig3}, respectively.
Figure \ref{Fig5b} shows the rotary-stage-type geometries combined with parabolic-type geometry
in the NA setup covering a narrow angle range from 9.3 deg.(left) to 24.8 deg.(right). 
The inducing beam and one of the two creation beams share the common parabolic mirror
fixed at the bottom optical table in the vacuum chamber.
The narrow angle incidence is realizable by changing the position of incidence of the inducing beam
on the surface of the common parabolic mirror by fixing the c1 creation beam.

\begin{figure}[]
\begin{center}
\includegraphics[keepaspectratio,scale=0.6]{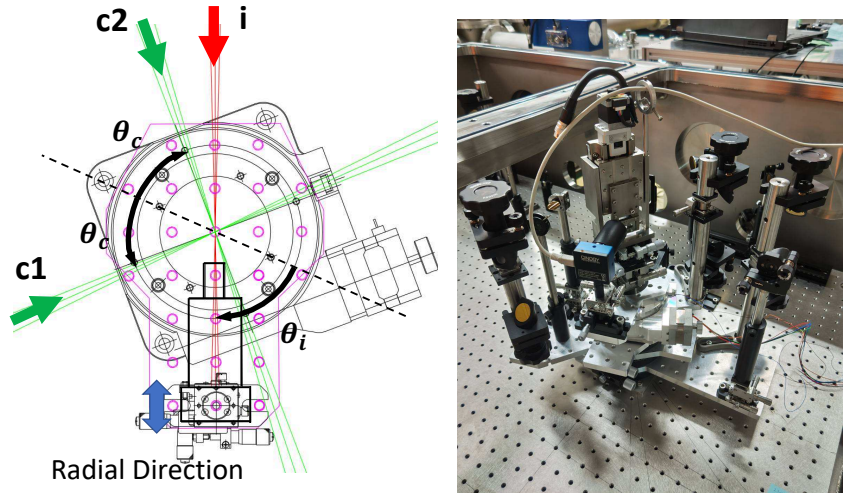}
\caption{Left: top view of the camera system on the top layer of the rotary stages to
monitor focal spots of all the three lasers c1, c2, and i.  Right: picture assembling
all the components for the large angle setup.
}
\label{Fig6}
\end{center}
\end{figure}

Figure \ref{Fig6} shows a top view of the top common camera layer 
together with a picture assembling all the components.
By preparing an independent layer only for the camera with a motorized rotation stage, 
beam profiles of all the three lasers at IP can be monitored and recorded. 
By reading the scale on the motorized rotation stage, the camera position can be adjusted 
to the incident angles $\pm \theta_{c}$ relative to the bisecting line (dashed line)
which is set by $-\theta_i$ with respect to the inducing beam direction. 
We note that the camera position must be calibrated so that the camera surface
is placed perpendicularly to the radial direction from the central IP position with a equal distance
for any rotated positions. By changing the camera position along the radial direction aligned to IP
by the local stage on which the camera is installed,
one can check whether a beam profile center stays at the same pixel point in the camera or not. 
If there is a drift of the profile center, that is, a deviation from the perpendicular direction, 
one can locally fine-tune the camera positions. 
With this fine-tuning method, AOIs can be adjusted with sufficient accuracy of 0.1\textdegree.

\section{Verification of rotary stage system}\label{sec4}
\begin{figure}[H]
\begin{center}
\includegraphics[keepaspectratio,width=125mm]{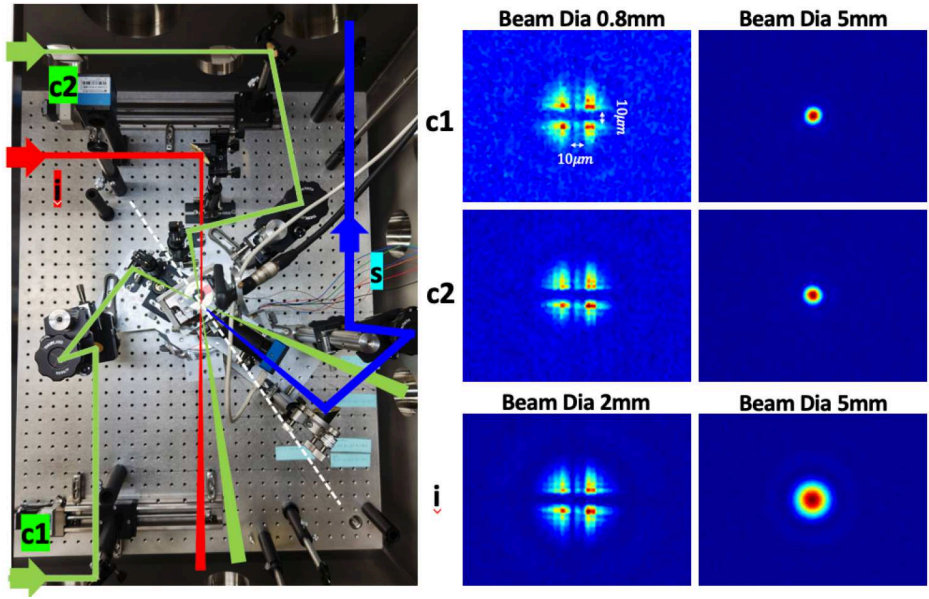}
\caption{Picture of large angle setup (left) and focal images of individual three beams (right)
when a common laser beam diameter of 5~mm are focused into IP
at $\theta_c=24.8$~degree. In the picture, the optical paths of the three beams, 
consisting of the creation beam (c1), the creation beam (c2), and the inducing beam (i), 
as well as the signal photon line (s) are drawn. 
The middle column shows the images of individual three lasers when they
hit the crossed point between two thin target wires of 10~$\mu$m diameter.
}
\label{Fig7}
\end{center}
\end{figure}

\begin{figure}[]
\begin{center}
\includegraphics[keepaspectratio,width=125mm]{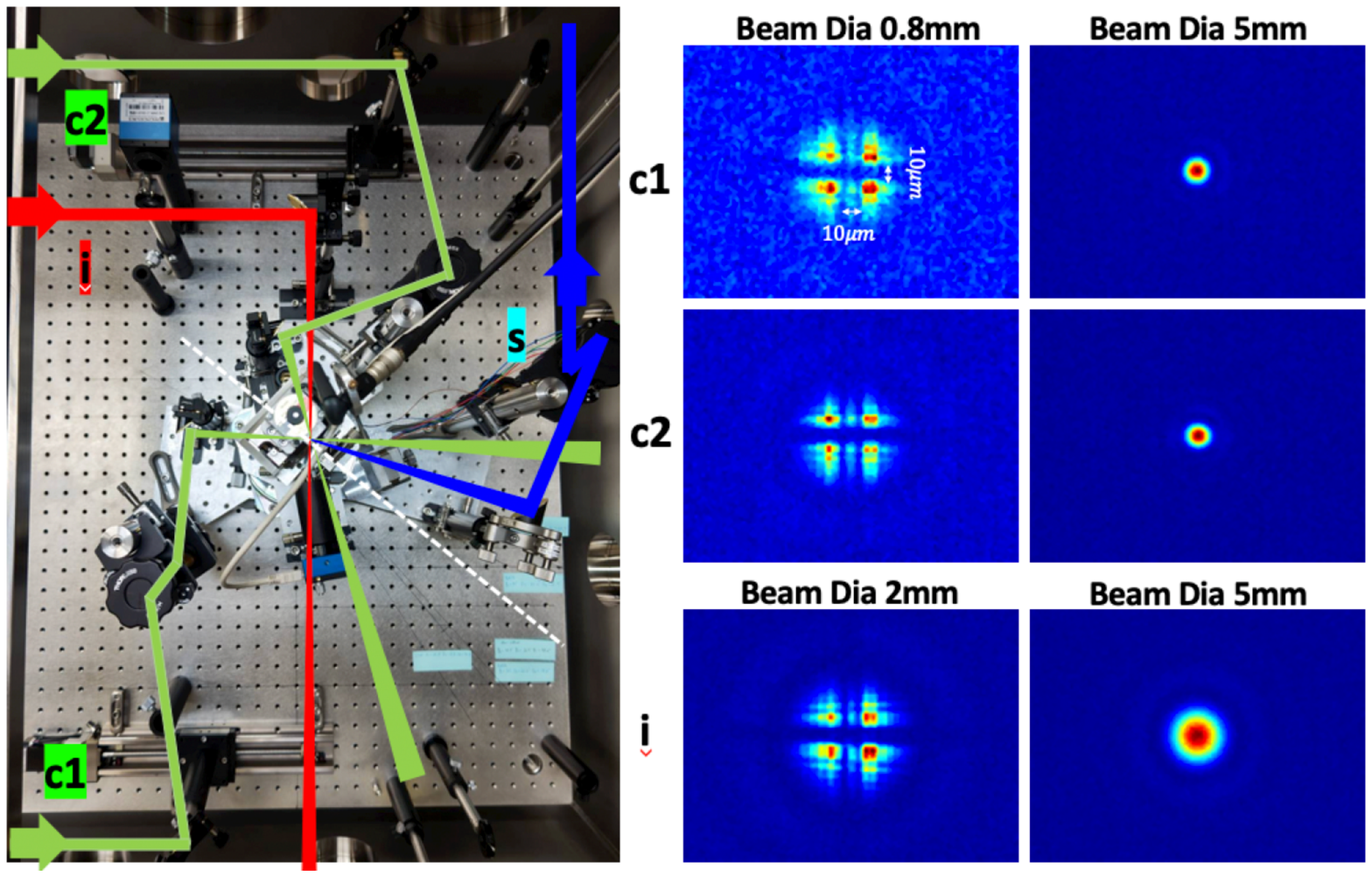}
\caption{
Picture of large angle setup (left) and focal images of individual three beams (right)
when a common laser beam diameter of 5~mm are focused into IP at $\theta_c=35.5$~degree. 
The other details are the same as in Fig \ref{Fig7}.} 
\label{Fig8}
\end{center}
\end{figure}

\begin{figure}[H]
\begin{center}
\includegraphics[keepaspectratio,width=125mm]{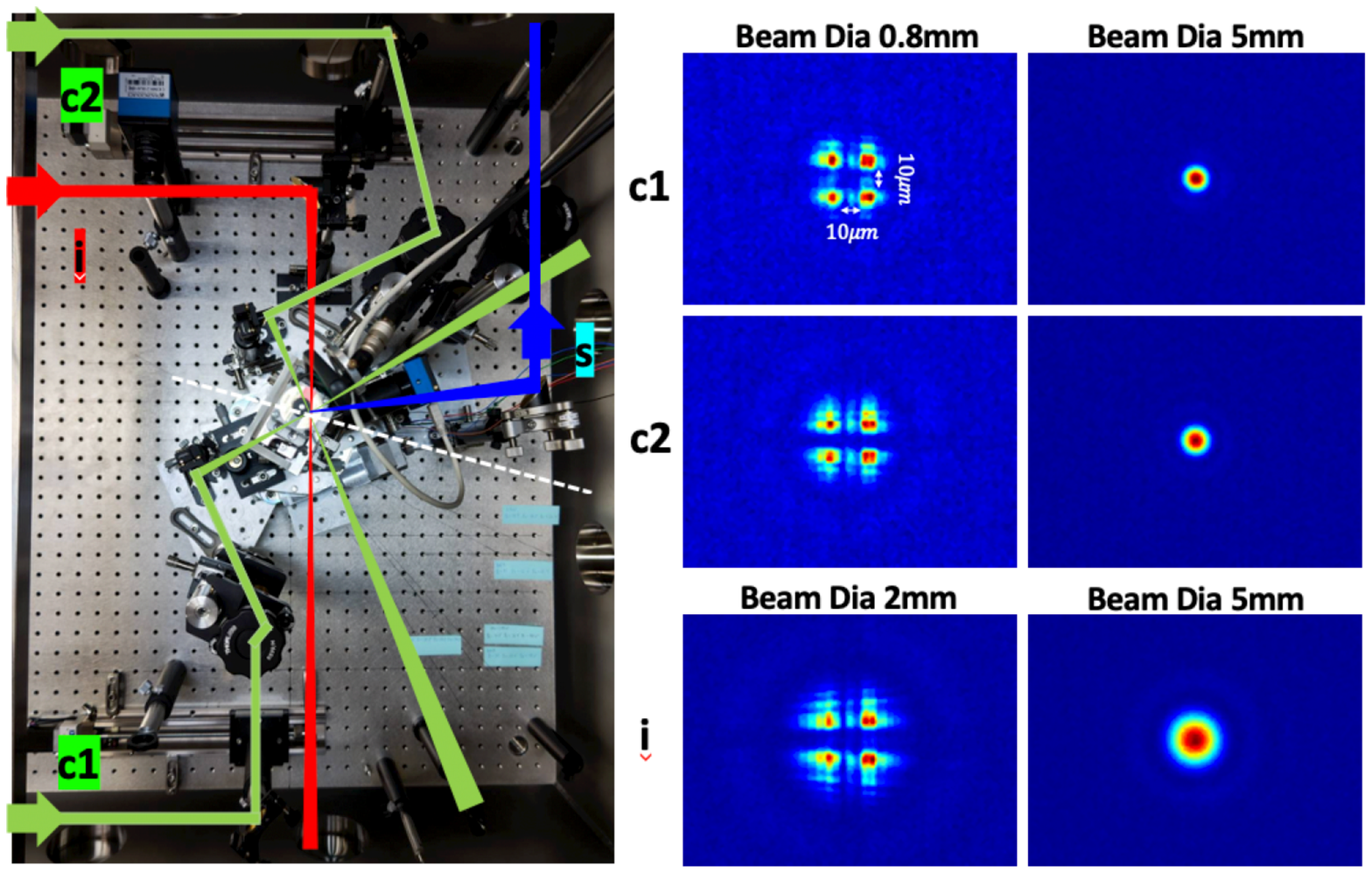}
\caption{
Picture of large angle setup (left) and focal images of individual three beams (right)
when a common laser beam diameter of 5~mm are focused into IP at $\theta_c=47.9$~degree. 
The other details are the same as in Fig \ref{Fig7}.} 
\label{Fig9}
\end{center}
\end{figure}

A three-beam SRPC covering for the LA collision case was actually constructed
after testing alignments of PS and PM on individual stages. 
Since the LA setup contains more rotary layers than the case of the NA setup, 
the verification of the LA setup guarantees the success of the NA case.
The spatial overlapping was then verified using a He-Neon laser.
Figure~\ref{Fig7}, \ref{Fig8} and \ref{Fig9} show the three collision geometries 
with incident angles of the creation beam at $\theta_c = 24.8$, 35.5, and 47.9 degrees, 
and the respective focal images of the three beams taken by a single camera are shown.
The optical paths of two creation beams, an inducing beam (c1, c2, i), and signal photons (s) 
are drawn for the reference. 
The focal images in the middle column show the spot profiles 
when the common 10~$\mu$m diameter crossed wire was placed
in front of the three beams (c1, c2, i) with a smaller beam diameter of 0.8~mm for the two creation beams
and 2~mm for the inducing beam to have broader focal images on purpose, 
while the focal images in the right column show the spot profiles at the same camera position 
after moving the target holder to
the position for the search mode (empty hole) by changing beam diameters to a common 5~mm 
which will be used for the future search.
We note exact focal lengths of the common creation beams and the inducing beam were 
101.6~mm 203.2~mm, respectively.

\section{Realistic sensitivity projections}
\begin{figure}[H]
\begin{center}
\includegraphics[keepaspectratio,scale=0.7]{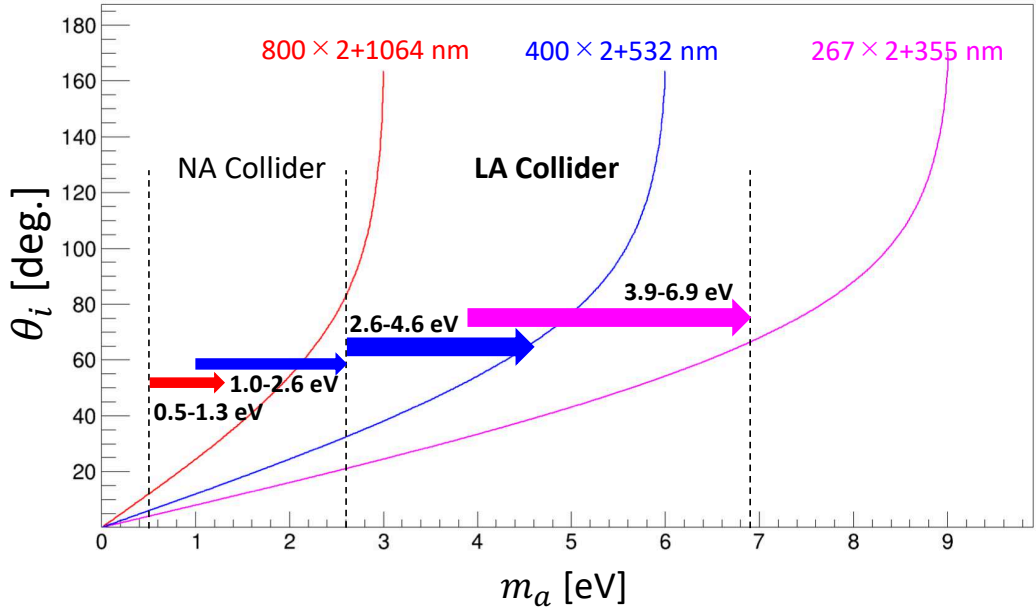}
\caption{
Incident angles of the inducing beam $\theta_i$ as a function of ALP mass $m_a$
which is determined by incident angles of creation beams $\theta_c$.
Three combinations of laser wavelengths for fundamental, second harmonic and third harmonic
cases. Namely, beginning with 800~nm (Ti:Sapphire) for two creation beams and 1064~nm (Nd:YAG)
for an inducing beam expressed as
$800 \times 2 + 1064$~nm (red), we extend the search to those with
$400 \times 2 + 532$~nm (blue) and $267 \times 2 + 355$~nm (magenta).
Depedning on the combinations between the two angle setups and laser wavelengths, accessible mass ranges
are different. This figure shows projections to cover from 0.5 to 6.9 eV.
}
\label{Fig10}
\end{center}
\end{figure}
We will provide sensitivity projections based on LA and NA collision setups in the following.
Figure \ref{Fig10} shows incident angles of the inducing beam $\theta_i$ as a function of ALP mass $m_a$
which is determined by incident angles of creation beams $\theta_c$ and their wavelengths.
We plan to use three combinations of laser wavelengths for fundamental, second harmonic and third harmonic
cases. Namely, beginning with 800~nm (Ti:Sapphire) for two creation beams and 1064~nm (Nd:YAG) 
for an inducing beam denoted as
$800 \times 2 + 1064$~nm (red), we extend the search to those with
$400 \times 2 + 532$~nm (blue) and $267 \times 2 + 355$~nm (magenta).
Depending on the combinations between the two angle setups and laser wavelengths, accessible mass ranges
are different. The figure shows our projections to cover from 0.5 to 6.9 eV 
with colored arrows corresponding to different wavelengths combinations 
where accessible mass ranges are specified.

\begin{figure}[H]
\begin{center}
\includegraphics[keepaspectratio,scale=0.7]{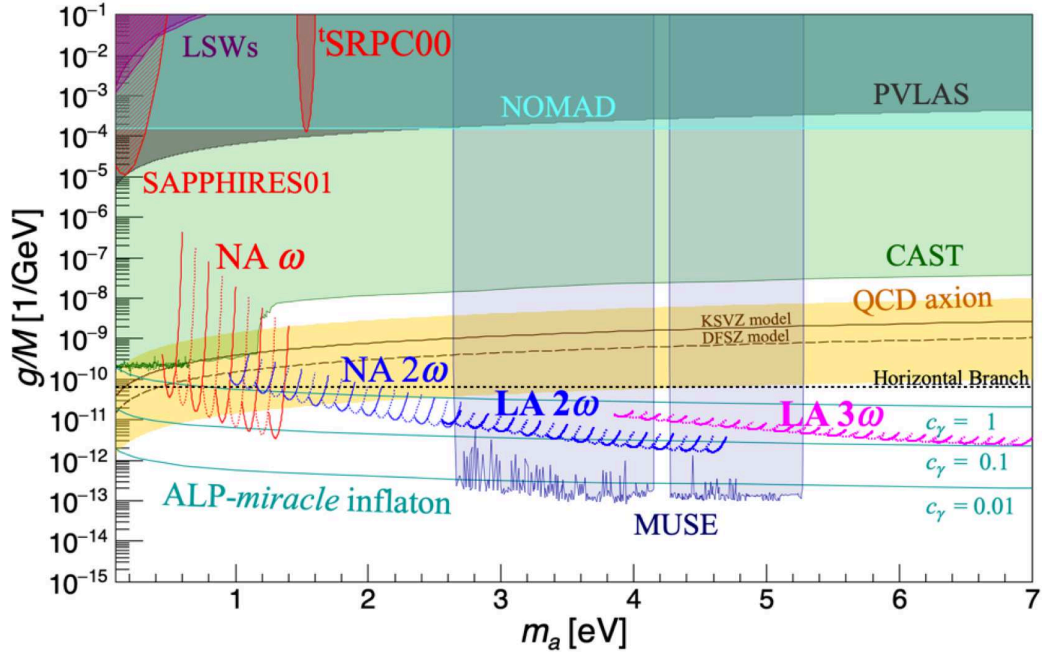}
\caption{
Sensitivity projections based on realistic parameters in Tab.\ref{Tab2}.
NA$\omega$, NA$2\omega$, LA$2\omega$ and LA$3\omega$ are sensitivities corresponding to
the four arrows in Fig.\ref{Fig10} specifying individual mass ranges.
The other details can be found in the main text.
} 
\label{Fig11}
\end{center}
\end{figure}

Accordingly, Fig.\ref{Fig11} shows the sensitivity projects based on parameters summarized in Tab.\ref{Tab2}.
Based on the set of realistic experimental parameters $P$ in Tab.\ref{Tab2},
the observed number of signal photons via an ALP exchange with the mass $m_{a}$ and the coupling $g/M$
to two photons is expressed as
        \begin{equation}
        \label{obtain_photon}
        n_{obs} = \mathcal{Y}_{c+i} \left(m_{a}, g / M; P\right) N_{shots} \epsilon,
        \end{equation}
where $\epsilon$ is the overall detection efficiency and
$N_{shots}$ is the number of shots.
A coupling constant $g /{M}$ can be numerically calculated by solving Eq.\eqref{obtain_photon}
for an ALP mass $m_{a}$ and a given observed number of photons $n_{obs}$.
We assumed the $n_{obs}$ as the noise originating photon-like signals $\delta N_{noise}$. 
NA$\omega$, NA$2\omega$, LA$2\omega$ and LA$3\omega$ are sensitivities corresponding to the individual arrows
classified in Fig.\ref{Fig10}.
The details of the numerical calculations and the derivation of the upper limits on the coupling
can be found in \cite{3beam00} and \cite{3beam01}, respectively.
The red shaded area shows the excluded range based on SRPC in
quasi-parallel collision geometry (SAPPHIRES01)~\cite{SAPPHIRES01}.
The red filled area indicates the excluded range with the fixed angle pilot search, 
\tSRPC00~\cite{3beam01}.
The gray area shows the excluded region by the vacuum magnetic birefringence experiment (PVLAS~\cite{PVLAS}).
The purple areas are excluded regions by the Light-Shining-through-a-Wall (LSW) experiments
(ALPS~\cite{ALPS} and OSQAR~\cite{OSQAR}).
The light-cyan horizontal solid line indicates the upper limit from the
search for eV (pseudo)scalar penetrating particles in the SPS neutrino beam (NOMAD)~\cite{NOMAD}.
The horizontal dotted line is the upper limit from the Horizontal Branch observation \cite{HB}.
The blue areas are exclusion regions from the optical MUSE-faint survey~\cite{MUSE}.
The green area indicates the excluded region by the helioscope experiment CAST \cite{CAST}.
The yellow band and the upper solid brown line are the predictions from the benchmark QCD axion models:
the KSVZ model \cite{KSVZ} with $0.07 < \left|E/N - 1.95\right| < 7$ and
$E/N = 0$, respectively, while the bottom dashed brown line is the prediction
from the DFSZ model \cite{DFSZ} with $E/N = 8/3$.
The cyan lines are the predictions from the ALP {\it miracle} model \cite{miracle}
with the model parameters $c_{\gamma}=1, 0.1, 0.01$.

\begin{table}[H]
\caption{
Experimental parameters used to numerically calculate the upper limits on the coupling--mass relations.
$(*)$ We note that the focal length of the inducing beam in the case of the narrow angle setup
must slightly vary in principle because of the nature of the parabolic mirror. 
However, since the incident position with respect to the focusing mirror does not vary a lot, 
for simplicity, we assume a common focal length to evaluate the sensitivity.
}
\begin{center}
\begin{tabular}{lr} \\ \hline
Parameters & Values\\ \hline
Two equal creation laser pulses \\ \hline
Central wavelength of creation laser $\lambda_c$ & 800~nm($\omega$)/400~nm($2\omega$)/267~nm(3$\omega$)\\
Relative linewidth of creation laser, $\delta\omega_c/<\omega_c>$ &  $10^{-2}$\\
Duration time of creation laser, $\tau_{c}$ & 40 fs \\
Creation laser energy per $\tau_{c}$, $E_{c}$ & 1~mJ \\
Beam diameter of creation laser beam, $d_{c}$ & 0.005~m\\
Focal length of narrow angle setup & $f_{c} = 0.18$~m \\
Focal length of large angle setup  & $f_{c} = 0.10$~m \\
Polarization & left-handed circular polarization \\ \hline
One inducing laser pulse \\ \hline
Central wavelength of inducing laser $\lambda_i$ & 1064~nm($\omega$)/532~nm($2\omega$)/355~nm(3$\omega$)\\
Relative linewidth of inducing laser, $\delta\omega_{i}/<\omega_{i}>$ &  $10^{-4}$\\
Duration time of inducing laser beam, $\tau_{i}$ & 9~ns\\
Inducing laser energy per $\tau_{i}$, $E_{i}$ & 100~mJ \\
Beam diameter of inducing laser beam, $d_{i}$ & $0.005$~m\\
Focal length of narrow angle setup $(*)$ & $f_{i} = 0.19$~m \\
Focal length of large angle setup        & $f_{i} = 0.20$~m \\
Polarization & right-handed circular polarization \\ \hline
Overall detection efficiency, $\epsilon$ & 5\% \\
Number of shots per collision angle, $N_{shots}$   & $10^4$ shots\\
$\delta{N}_{noise}$ & 50\\
\hline
\end{tabular}
\end{center}
\label{Tab2} 
\end{table}

\section{Conclusions and Future Plans}
We have designed two types of variable-angle stimulated resonant photon colliders 
with three laser beams (\tSRPC) covering narrow and large angles, respectively.
The large angle setup sensitive to relatively a higher mass range was actually constructed,
and the mechanism was verified using a He-Neon laser for the calibration.
We confirmed that the incident angle can be varied by using a rotating stage and a periscope, 
and we ensured the spatial overlapping of three beam focal spots
at multiple collision angles by developing a monitoring system 
that allows a single camera to check the focal spot images.
As in the previous pilot search at the fixed incident angle~\cite{3beam01}, 
time synchronization is expected to be ensured by switching the focal point target
to a nonlinear crystal, BBO crystal, and using a delay line with a retro-reflector 
when a high intensity laser is used.

Given the realistic designs for both narrow and large angle setups, we have provided sensitivity
projections in the near future searches for ALPs based on \tSRPC~with possible
combinations of three laser wavelengths. 
The sensitivity projects show that the proposed collider can reach coupling domains 
relevant to the QCD axion models and the {\it Miracle} scenario 
over the mass range of 0.5 - 6.9 eV within the present reach of laser technologies.

\section*{Acknowledgments}
K. Homma acknowledges the support of the Collaborative Research
Program of the Institute for Chemical Research at Kyoto University 
(Grant Nos.\  2023--101) and Grants-in-Aid for Scientific Research
Nos.\ 21H04474 from the Ministry of Education, 
Culture, Sports, Science and Technology (MEXT) of Japan.
Y. Kirita acknowledges support from a Grant-in-Aid for JSPS fellows No. 22J13756 from the Ministry of Education, Culture, Sports, Science and Technology (MEXT) of Japan.
%


\end{document}